# Spin-current-induced charge current

Soon-wook Jung and Hyun-Woo Lee

*Department of Physics, Pohang University of Science and Technology, Pohang, Kyungbuk, 790-784, Korea*

**Abstract**

We show that the injection of a *pure* spin current (not accompanied by charge current) into a ring can induce a circulating charge current in the ring, provided that transport coefficients of the ring are spin-dependent *and* inhomogeneous. As an example, we consider a hybrid ferromagnet(F)-normal metal(N) ring system and calculate the magnitude of the charge current induced by the pure spin current injection. This phenomenon may have relevance for spintronic applications.

Spintronics aims to develop methods to utilize the electron's spin degree of freedom [1]. Successful utilization requires the development of various techniques such as generation of spin-polarized current, manipulation of spin, spin relaxation control, and the detection of spins. Recently the conversion from electronic signal/degree of freedom to spin current becomes of interest [1] in view of their possible relevance for spintronic applications, especially for the generation of a spin-polarized current. Some proposed conversion mechanisms [2] exploit properties such as the Coulomb blockade of quantum dots [3] and/or the coherence of electron phase [3]. Very recently an interesting possibility of spin current generation via spin Hall effect [4,5] is also proposed. Furthermore, the inverse conversion from the spin signal to electric signal is also of interest [6,7]. It is demonstrated that due to the relativistic covariance of Maxwell equations, a steady-state pure spin current can induce an electric field [7]. Such inverse conversion may be useful for spin current detection.

In this paper, we will study diffusion dynamics of electrons in a ring and show that it can result in spin-current-induced charge current, namely, charge current in a diffusive ring induced by the injection of a pure spin current (pure in the sense that it is *not* accompanied by charge injection) into the ring. Although the phenomenon is similar to that in Ref. [7], the mechanism that we will deal with is completely different from that in Ref. [7]. Due to this difference, this phenomenon can occur even when relativistic effects vanish or are negligible. An example illustrating this point will be discussed below. We find that the diffusion dynamics can give rise to finite charge current from the injected pure spin current when the following two requirements are satisfied: (A) electron transport coefficients (such as electric conductivity) of the ring are spin dependent *and* (B) the ring is inhomogeneous (for example, position dependent electric conductivity).

The requirement (A) is intuitively appealing; Without any spin dependencies, spin-up and spin-down currents will be the same in magnitude and opposite in direction, so that the resulting charge current vanishes (see Appendix A for proof in the linear response regime). The reason for the requirement (B) may be less evident. To illustrate the reason, consider a one-dimensional ring. When the pure spin current injection rate is kept at a constant value, the system reaches a steady state and the steady state charge current $I_e = I^\uparrow + I^\downarrow$ is homogeneous over the ring due to the charge conservation $\partial \rho_e / \partial t + (1/S) \partial I_e / \partial x = 0$ where $\partial \rho_e / \partial t = 0$ in the steady state and $S$ is the cross-sectional area of the wire forming the ring. Then $I_e$ may be re-expressed as follows,

$$I_e = \frac{1}{L}\oint_{\text{ring}} dx\, I_e(x) = -\frac{1}{L}\oint_{\text{ring}} dx\, \frac{S}{e}\left(\sigma^\uparrow \frac{\partial \mu^\uparrow}{\partial x} + \sigma^\downarrow \frac{\partial \mu^\downarrow}{\partial x}\right),$$

where $\sigma^{\uparrow(\downarrow)}$ is the electric conductivity, $\mu^{\uparrow(\downarrow)}$ is the electrochemical potential for spin $\uparrow(\downarrow)$, and $L$ is the circumference of the ring. Now if the ring is homogeneous and thus both $\sigma^\uparrow$ and $\sigma^\downarrow$ are homogeneous (independent of $x$), the last expression vanishes identically since $\mu^\uparrow$ and $\mu^\downarrow$ are single-valued functions, $\mu^{\uparrow(\downarrow)}(x=0) = \mu^{\uparrow(\downarrow)}(x=L)$. This simple analysis illustrates the origin of the requirement (B).

To demonstrate further the requirements (A) and (B) and also to illustrate the difference from Ref. [7], we examine a particular example of a one-dimensional ring (Fig. 1) of length $L$ made of a ferromagnet (F) with length $L_F$ and a normal metal (N) with length $L_N$ ($L = L_F + L_N$). This hybrid ring system satisfies the both requirements for the conversion phenomenon. For definiteness, we assume that the F is magnetized along the $-y$-axis and that the injected pure spin current $I_s \equiv I_0^\uparrow - I_0^\downarrow$ is polarized along the same axis. Here $I_0^\uparrow$ ($I_0^\downarrow$) is the injected spin-up (spin-down) current into the ring and we set their magnitude to be the same but their direction opposite so that the injected net charge current vanishes. There are various ways to realize such pure spin current [2, 8-10]. One particular example will be considered in Appendix B. Due to the requirement (B), the conversion is expected to be more efficient when the spin current is injected to the region where the conductivities change, namely, FN interfaces. Thus we consider the pure spin current injected to the FN interface I-1 ($l=0$ in Fig. 1). We also assume that the FN interfaces (both I-1 and I-2) are ohmic. We remark that in this configuration, where the spin current direction lies in the $xy$-plane, the electric field induced by the injected spin current (due to the mechanism in [7]) is perpendicular (see Eq. (9) in Ref. [7]) to the ring plane and thus cannot induce a circulating charge current in the ring (actually the induced electric field due to the mechanism in [7] is zero in Fig. 1 since the spin current direction is parallel to its spin polarization direction). Therefore the induced charge current in this example has clearly nothing to do with the mechanism in Ref. [7].

In order to study the response of the diffusive system to the pure spin injection, we use a simple spin diffusion model [11-15], which describes the electrical transport in terms of two current channels (spin-up and spin-down current channels). The conductivity of

each channel is given by

$$\sigma^{\uparrow(\downarrow)} = e^2 N^{\uparrow(\downarrow)} D^{\uparrow(\downarrow)}, \tag{1}$$

where $N^{\uparrow(\downarrow)}$ denotes the spin dependent density of states at the Fermi energy, $D^{\uparrow(\downarrow)}$ is the spin dependent diffusion constant, and e(<0) is the electron charge. Due to the spin dependent conductivities, the current in the ferromagnet is spin-polarized with the polarization given by

$$\alpha = (\sigma^{\uparrow} - \sigma^{\downarrow})/(\sigma^{\uparrow} + \sigma^{\downarrow}). \tag{2}$$

Spin flip processes can be characterized by the spin flip time $\tau^{\uparrow\downarrow}$, the average time to flip an up-spin to a down-spin, and $\tau^{\downarrow\uparrow}$ for the reverse process. Using the detailed balance principle $N^{\uparrow}/\tau^{\uparrow\downarrow} = N^{\downarrow}/\tau^{\downarrow\uparrow}$ for equilibrium states with no net spin flip, one obtains the equations for spin transport:

$$j^{\uparrow(\downarrow)} = -\frac{\sigma^{\uparrow(\downarrow)}}{e}\frac{\partial \mu^{\uparrow(\downarrow)}}{\partial x}, \tag{3}$$

$$\frac{\partial^2 (\mu^{\uparrow} - \mu^{\downarrow})}{\partial x^2} = \frac{(\mu^{\uparrow} - \mu^{\downarrow})}{\lambda}, \tag{4}$$

where $\lambda = \sqrt{D\tau}$ is the spin diffusion length, $D = D^{\uparrow}D^{\downarrow}(N^{\uparrow}+N^{\downarrow})/(N^{\uparrow}D^{\uparrow}+N^{\downarrow}D^{\downarrow})$ is the average diffusion constant, $\tau = \tau^{\uparrow\downarrow} \cdot \tau^{\downarrow\uparrow}/(\tau^{\uparrow\downarrow} + \tau^{\downarrow\uparrow})$ is the average spin relaxation time, and $\mu^{\uparrow(\downarrow)}$ is the spin dependent electrochemical potential (ECP). For the convenience of the solution construction, we split the system into two regions, PI ($0 < l < L_N$) and FI ($L_N < l < L_N + L_F = L$ or $0 < l' < L_F$ where $l' \equiv L - l$), see Fig. 1. Then the ECPs $\mu_F^{\uparrow(\downarrow)}$ in the region FI and $\mu_N^{\uparrow(\downarrow)}$ in the region PI are given by

$$\mu_F^{\uparrow(\downarrow)} = A + B \cdot l' \pm \frac{1}{\sigma_F^{\uparrow(\downarrow)}}\left[C\exp\left(-\frac{l'}{\lambda_F}\right) + D\exp\left(\frac{l'-L_F}{\lambda_F}\right)\right], \tag{5}$$

$$\mu_N^{\uparrow(\downarrow)} = E \cdot l \pm \left[F\exp\left(-\frac{l}{\lambda_N}\right) + G\exp\left(\frac{l-L_N}{\lambda_N}\right)\right], \tag{6}$$

where the seven constants $A$, $B$, …, $G$ should be chosen to satisfy the ohmic boundary conditions, namely, the continuity of $\mu^{\uparrow(\downarrow)}$ and the current conservation for each spin channel at the interfaces.

After some algebra, the charge current $I_e = I^\uparrow + I^\downarrow$ induced in the FN hybrid ring due to the injected pure spin current $I_s$ is found to be

$$I_e = I_s \times \frac{\dfrac{\alpha}{1-\alpha^2}\dfrac{R_F}{R_N}(e^{\zeta_F}-1)(e^{\zeta_N}-1)}{\dfrac{R_{\text{ring}}^{\text{ohm}}}{R_N}\left[(e^{\zeta_F}+1)(e^{\zeta_N}-1) + \dfrac{1}{1-\alpha^2}\dfrac{R_F}{R_N}(e^{\zeta_F}-1)(e^{\zeta_N}+1)\right] + \dfrac{2\alpha^2}{1-\alpha^2}\dfrac{R_F}{R_N}(e^{\zeta_F}-1)(e^{\zeta_N}-1)}, \quad (7)$$

where $R_{\text{ring}}^{\text{ohm}} \equiv S(L_F/\sigma_F + L_N/\sigma_N)$ is the resistance of the ring, $R_F \equiv S\lambda_F/\sigma_F$ and $R_N \equiv S\lambda_N/\sigma_N$ are the resistances corresponding to the spin diffusion lengths $\lambda_F$ and $\lambda_N$, respectively, and $\zeta_F \equiv L_F/\lambda_F$, $\zeta_N \equiv L_N/\lambda_N$ are, respectively, the dimensionless lengths of the F and N parts normalized with respect to the corresponding spin diffusion lengths.

A few limiting cases deserve further discussion. Note that $I_e$ vanishes when $\zeta_N \to 0$ or $\zeta_F \to 0$ (homogeneous ring limit). Thus when the transport properties becomes homogeneous, the charge current is not induced any more, illustrating the necessity of the requirement (B). Note also that $I_e$ also vanishes when $\alpha \to 0$, regardless of the ratio $\sigma_F/\sigma_N$. Thus when the ring is inhomogeneous ($\sigma_F \neq \sigma_N$) but the transport has no spin dependencies ($\alpha = 0$), the charge current is not induced, illustrating the necessity of the requirement (A). When none of above applies, that is, when $\zeta_N \neq 0$, $\zeta_F \neq 0$, and $\alpha \neq 0$, the induced charge current has a nonzero value. Thus Eq. (7) confirms that the pure spin current $I_s$ injected into the FN hybrid structure indeed generates a nonzero charge current $I_e$. In Eq. (7), the sign convention of $I_e$ is chosen in such a way that the charge current is positive if it flows in the counterclockwise direction. Thus for $I_s > 0$, the charge current flows in the counterclockwise/clockwise direction when the magnetization direction of the F is parallel ($\alpha > 0$)/anti-parallel ($\alpha < 0$) to the polarization of the injected spin current.

For a large ring with $\zeta_F, \zeta_N \gg 1$ (so that $R_{\text{ring}}^{\text{ohm}} \gg R_F, R_N$), Eq. (7) is considerably simplified to

$$I_e R_{\text{ring}}^{\text{ohm}} = I_s \cdot \frac{\alpha R_F}{1-\alpha^2} \frac{R_N}{R_N + \dfrac{R_F}{1-\alpha^2}}. \quad (8)$$

Recalling that $R_{ring}^{ohm}$ is the resistance of the ring, the right-hand-side can be interpreted as the electromotive force (EMF) induced by the pure spin current injection. When there is a serious conductance mismatch problem, $R_N \gg R_F/(1-\alpha^2)$ [14-16], the induced electromotive force reduces further to $I_s \cdot \alpha R_F/(1-\alpha^2)$, which depends only on properties of the F side, the part with smaller resistance. Note that in Eq. (8), the length dependence appears in $R_{ring}^{ohm}$ only and roughly speaking, $I_e$ decays inversely proportional to the ring circumference. Thus even though the injected spin current decays exponentially fast due to spin relaxation and hardly propagates in the ring with $\zeta_F, \zeta_N \gg 1$, it still manages to induce a long-ranged effect that decays with a power law ($1/L$).

Here we present an interesting identity that gives insight into the origin of the spin-current-induced charge current. One first defines "weighted" voltage differences $V_{I-1}$ and $V_{I-2}$ at the two particular positions where the transport coefficients change;

$$V_{I-1} \equiv \frac{1}{e}\left[\left(\frac{1+\alpha}{2}\mu_F^\uparrow + \frac{1-\alpha}{2}\mu_F^\downarrow\right)\bigg|_{l=L^-} - \left(\frac{\mu_N^\uparrow + \mu_N^\downarrow}{2}\right)\bigg|_{l=0^+}\right], \tag{9}$$

$$V_{I-2} \equiv \frac{1}{e}\left[\left(\frac{1+\alpha}{2}\mu_F^\uparrow + \frac{1-\alpha}{2}\mu_F^\downarrow\right)\bigg|_{l=L_N^+} - \left(\frac{\mu_N^\uparrow + \mu_N^\downarrow}{2}\right)\bigg|_{l=L_N^-}\right], \tag{10}$$

which reduce to usual voltage differences if there are no spin dependencies. Note that due to the spin dependent conductivities ($\alpha \neq 0$), $V_{I-1}$ and $V_{I-2}$ can have nonzero values despite the continuity condition $\mu_F^{\uparrow(\downarrow)} = \mu_N^{\uparrow(\downarrow)}$ at the interfaces. We find that regardless of values of various parameters, the following relation holds,

$$I_e \cdot R_{ring}^{ohm} + V_{I-2} = V_{I-1}, \tag{11}$$

which indicates that the voltage $V_{I-1}$ at the spin injection point can be interpreted as an effective EMF while the voltage $V_{I-2}$ as a voltage drop at I-2. In the large ring limit $\zeta_F, \zeta_N \gg 1$, a further insight can be obtained by considering the following deformation to Fig. 1; Imagine cutting the connection at the interface I-2 (as shown in Fig. 2) and setting the ECP's at $x = \infty$ and $y = \infty$ so that no charge current flows in the normal and ferromagnet. The deformation from Fig.1 to Fig.2 affects the profiles of the ECP's; While the ECP's near the interface I-1 ($x = 0$, $y = 0$) is not modified since the I-1 is infinitely away from the I-2, the ECP's near the I-2 are modified and "locked" to the weighted averages of the ECP's near the I-1 as follows,

$$\mu_N^\uparrow(x = \infty) = \mu_N^\downarrow(x = \infty) = \mu_N^{\text{ave}}(x = 0^+) \equiv \left( \frac{\mu_N^\uparrow + \mu_N^\downarrow}{2} \right)\bigg|_{x=0^+}, \qquad (12)$$

$$\mu_F^\uparrow(y = \infty) = \mu_F^\downarrow(y = \infty) = \mu_F^{\text{ave}}(y = 0^+) \equiv \left( \frac{1+\alpha}{2} \mu_F^\uparrow + \frac{1-\alpha}{2} \mu_F^\downarrow \right)\bigg|_{y=0^+}, \qquad (13)$$

due to the condition of no charge current. Note that the ECP's at the I-2 is not continuous any more, $\mu_N^\uparrow(x = \infty) \neq \mu_F^\uparrow(y = \infty)$ and $\mu_N^\downarrow(x = \infty) \neq \mu_F^\downarrow(y = \infty)$. Thus the condition of the no charge current is *not compatible* with the continuity of the ECP's. Therefore when the continuity of the ECP's at the I-2 is enforced as in Fig. 1, the charge current should appear. We remark that the voltage difference $V(y = \infty) - V(x = \infty) = \left[ \mu_F^{\text{ave}}(y = \infty) - \mu_N^{\text{ave}}(x = \infty) \right]/e$ for the configuration in Fig. 2 is proportional to the spin accumulation $\mu_F^\uparrow(y = 0) - \mu_F^\downarrow(y = 0) = \mu_N^\uparrow(x = 0) - \mu_N^\downarrow(x = 0)$ near the I-1 and often called the spin EMF. The spin EMF has been demonstrated to be a useful detection tool of the spin accumulation [17]. We also remark that the spin EMF for the configuration in Fig. 2 is identical to the EMF in the large ring limit [right-hand-side in Eq. (8)] for the configuration in Fig. 1. In this sense the spin EMF may be regarded as a limiting case of the spin-current-induced charge current.

Next we lift the assumption of the ohmic interfaces and consider interfaces with tunneling barriers to show that the ohmic interface is not essential for nonvanishing charge current. Since the ECPs $\mu^\uparrow$ and $\mu^\downarrow$ are now discontinuous at the interfaces, it is necessary to specify the spin injection position more precisely. For definiteness, we consider the spin current injected to the normal metal side of the I-1 interface, that is, to $l = 0^+$. For simplicity, we assume that the tunneling barriers are nonmagnetic and thus there is no spin flip while electrons tunnel the tunneling barriers. Effects of the tunneling barriers can be described in terms of interface conductances $G_i^{\uparrow(\downarrow)}$

[12,13,18,19], which relate the discontinuities in the ECPs with the tunneling currents $I_i^{\uparrow(\downarrow)}$ through the interface I-1,

$$I_i^{\uparrow} = -\frac{G_i^{\uparrow}}{e}(\mu_F^{\uparrow}|_{l=0^+} - \mu_N^{\uparrow}|_{l=0^-})$$
$$I_i^{\downarrow} = -\frac{G_i^{\downarrow}}{e}(\mu_F^{\downarrow}|_{l=0^+} - \mu_N^{\downarrow}|_{l=0^-})$$
(14)

where $l = 0^{+(-)}$ represents the location of the I-1 approached from the N (F) side. $G_i^{\uparrow}$ and $G_i^{\downarrow}$ are in general different even for nonmagnetic tunneling barriers [19] since they depend on the densities of states of the F, which usually depend on spin direction. Here the tunnel barrier resistance $R_i$ is defined as $G_i = G_i^{\uparrow} + G_i^{\downarrow} = R_i^{-1}$. Similar relations are assumed for the I-2 as well. After some algebra, one obtains the charge current,

$$I_e = I_s \cdot \frac{\frac{p}{1-p^2}\frac{R_i}{R_N}(e^{\zeta_F}+1)(e^{\zeta_N}-1) + \frac{\alpha}{1-\alpha^2}\frac{R_F}{R_N}(e^{\zeta_F}-1)(e^{\zeta_N}-1)}{\frac{R_{\text{ring}}^{\text{ohm}}}{R_N}C_1 + \frac{2}{1-p^2}\frac{R_i}{R_N}C_2 + \frac{2\alpha^2}{1-\alpha^2}\frac{R_F}{R_N}(e^{\zeta_F}-1)(e^{\zeta_N}-1)},$$
(15)

where the dimensionless constants $C_1$ and $C_2$ are given by

$$C_1 = (e^{\zeta_F}+1)(e^{\zeta_N}-1) + \frac{1}{1-\alpha^2}\frac{R_F}{R_N}(e^{\zeta_F}-1)(e^{\zeta_N}+1) + \frac{1}{1-p^2}\frac{R_i}{R_N}(e^{\zeta_F}+1)(e^{\zeta_N}+1),$$

$$C_2 = (e^{\zeta_F}+1)(e^{\zeta_N}-1) + \frac{(1-2p\alpha+\alpha^2)}{1-\alpha^2}\frac{R_F}{R_N}(e^{\zeta_F}-1)(e^{\zeta_N}+1) + \frac{R_i}{R_N}(e^{\zeta_F}+1)(e^{\zeta_N}+1).$$

Here $p \equiv (G_i^{\uparrow} - G_i^{\downarrow})/(G_i^{\uparrow} + G_i^{\downarrow})$ is the spin polarization of the interfacial current. Note that when $\zeta_N \to 0$, the charge current vanishes since the requirement (B) is not satisfied. When $\zeta_F \to 0$, on the other hand, the charge current does *not* necessarily vanish, unlike the ohmic interface case. This difference arises from the fact that the F region of very small width can still endow a finite spin polarization $p \neq 0$ for the interfacial current. In the physical limit where the ferromagnetic region is totally removed from the ring, both $\varsigma_F$ and $p$ should approach zero simultaneously. Then the requirement (B) is not satisfied and the induced charge current vanishes. Similarly $\alpha \to 0$ alone does not guarantee the zero current; it should be combined with $p \to 0$ to make the requirement (A) not satisfied and thus make the charge current vanish. When none of these conditions for the zero charge current applies, the charge current is nonzero. Thus Eq. (15) confirms the existence of a nonzero charge current even in the presence of tunneling barriers.

We examine a few limiting cases. In the weak tunneling barrier limit $R_i \to 0$, Eq. (15) reduces to the ohmic result [Eq. (7)], as it should. In the strong tunneling barrier limit, $R_i \gg R_N, R_F$, Eq. (15) is simplified to

$$I_e \cdot R_{ring}^{tunnel} = I_s \cdot p R_N \tanh \frac{\varsigma_N}{2}, \tag{16}$$

where $R_{ring}^{tunnel} = R_{ring}^{ohm} + 2R_i$ is the resistance of the ring with two tunnel barriers. Thus the right-hand-side of Eq.(16) can be interpreted as the electromotive force induced by the pure spin current injection to the ring with strongly tunneling-barriered interfaces.

We next consider the large ring limit, $\varsigma_N \to \infty$, $\varsigma_F \to \infty$ ($R_{ring}^{tunnel} \gg R_F$, $R_N$, $R_i$). In this limit, $I_e$ obviously approaches zero due to the diverging resistance $R_{ring}^{tunnel}$. The electromotive force $I_e \cdot R_{ring}^{tunnel}$ on the other hand approaches a nonzero finite value,

$$I_e \cdot R_{ring}^{tunnel} \to I_s R_N \cdot \frac{\dfrac{p}{1-p^2}\dfrac{R_i}{R_N} + \dfrac{\alpha}{1-\alpha^2}\dfrac{R_F}{R_N}}{1 + \dfrac{1}{1-p^2}\dfrac{R_i}{R_N} + \dfrac{1}{1-\alpha^2}\dfrac{R_F}{R_N}}. \tag{17}$$

This value is again the same as the spin EMF for the configuration in Fig. 2 if the junction in Fig.2 is not ohmic but characterized by the same tunneling barrier parameters $p$ and $R_i$. Thus even for junctions with tunneling barriers, the spin EMF can be related to the large ring limiting case of the spin-current-induced charge current. This interesting connection with the spin EMF suggests that the phenomenon of the charge current induction by the spin current may also be useful for spin current detection.

Finally we estimate the magnitude of the ratio $I_e / I_s$ for $\alpha = 0.7$, $p = 0.4$. When a ferromagnetic metal of $L_F = 0.1\lambda_N$ and a normal metal of $L_N = 2\lambda_N$ form a ring with ohmic interfaces, Eq. (7) results in $I_e/I_s \sim 0.035$ for $\sigma_F \sim 10^7 \Omega^{-1} m^{-1}$, $\lambda_F \sim 10 nm$, $\sigma_N \sim 10^8 \Omega^{-1} m^{-1}$, $\lambda_N \sim 1 \mu m$ [14]. Thus the induced charge current $I_e$ is smaller than $I_s$ but not unmeasurably small. On the other hand, in tunneling-barriered interfaces with $R_i \sim 10^3 \Omega$, Eq. (16) results in $I_e/I_s \sim 0.0003$, which is much smaller than that for the ohmic interface case. When the normal metal is replaced by a semiconductor, $I_e/I_s$ for the tunneling-barriered interfaces can increase by about two orders of magnitude because the resistivity of a semiconductor can be larger than that of a normal metal by factor $10^2$.

To summarize, we demonstrated that a pure spin current injected into a diffusive ring can induce a circulating charge current provided that the two requirements ((A)spin dependent transport coefficients and (B)inhomogeneity) are satisfied. As a particular example, an F-N hybrid ring has been analyzed in detail to demonstrate various features of the induced charge current. An interesting connection with the spin EMF was also pointed out. It was also noted that despite close similarity with the phenomenon of the spin-current-induced electric field [6,7], the phenomenon that we addressed has a different origin and arises instead from spin-dependent diffusion dynamics. The example of the F-N ring demonstrates clearly this difference. This phenomenon of the spin-current-induced charge current may be useful for spintronic applications.

This work was initiated during the 2003 APCTP Focus Program. This work was supported by the SCORE-A and the electron Spin Science Center funded by the Korea Science and Engineering Foundation, and the Nano Research and Development Program funded by the Ministry of Science and Technology in Korea.

**Appendix A: No charge current induction without spin dependencies**

Here we show that at least in the linear regime, an injected spin current cannot induce a circulating charge current in a diffusive ring unless transport properties of the ring are spin dependent. One begins with the diffusion equations for the ECP's:

$$N^\uparrow \frac{\partial \mu^\uparrow}{\partial t} = \frac{\partial}{\partial x}\left(\frac{\sigma^\uparrow}{e}\frac{\partial \mu^\uparrow}{\partial x}\right) - \frac{N^\uparrow}{\tau^{\uparrow\downarrow}}\mu^\uparrow + \frac{N^\downarrow}{\tau^{\downarrow\uparrow}}\mu^\downarrow, \tag{A.1}$$

$$N^\downarrow \frac{\partial \mu^\downarrow}{\partial t} = \frac{\partial}{\partial x}\left(\frac{\sigma^\downarrow}{e}\frac{\partial \mu^\downarrow}{\partial x}\right) - \frac{N^\downarrow}{\tau^{\downarrow\uparrow}}\mu^\downarrow + \frac{N^\uparrow}{\tau^{\uparrow\downarrow}}\mu^\uparrow, \qquad (A.2)$$

where $N^\uparrow$ and $N^\downarrow$ are densities of states at the Fermi energy for spin up and down electrons, respectively. If transport properties are independent of spin, that is, $N^\uparrow = N^\downarrow = N_0$, $\sigma^\uparrow = \sigma^\downarrow = \sigma_0$, $\tau^{\uparrow\downarrow} = \tau^{\downarrow\uparrow} = \tau_0$, one obtains

$$N_0 \frac{\partial}{\partial t}\left(\mu^\uparrow + \mu^\downarrow\right) = \frac{\partial}{\partial x}\left[\frac{\sigma_0}{e}\frac{\partial}{\partial x}\left(\mu^\uparrow + \mu^\downarrow\right)\right], \qquad (A.3)$$

by summing up Eqs. (A.1) and (A.2). In the steady state, the left-hand-side vanishes by definition and the right-hand-side should vanish as well. By integrating the right-hand-side, one obtains

$$\mu^\uparrow(x) + \mu^\downarrow(x) = A_1 e \cdot \int_0^x dx \frac{1}{\sigma_0(x)} + A_2, \qquad (A.4)$$

where $A_1$ and $A_2$ are integration constants. To be consistent with the periodic boundary condition $\mu^\uparrow(x=L) + \mu^\downarrow(x=L) = \mu^\uparrow(x=0) + \mu^\downarrow(x=0)$, $A_1$ should be set to zero. Otherwise $\mu_\uparrow(x) + \mu_\downarrow(x)$ becomes a monotonically increasing (decreasing) function of $x$ depending on the sign of $A_1$. Now recalling the relation,

$$j^\uparrow + j^\downarrow = -\frac{\sigma_0}{e} \cdot \frac{\partial}{\partial x}\left(\mu^\uparrow + \mu^\downarrow\right), \qquad (A.5)$$

we verify that the circulating charge current should indeed vanish if transport properties are spin-independent.

**Appendix B: An example of a pure spin current injection**

As a particular example of a pure spin current injection method, we consider geometry in Fig. 3, where the normal metal segment $P_0$ connects the ring to the path $P_s$ on which a charge current $I$ flows. Note that despite the connection, the charge current cannot flow into the ring due to the charge conservation of the ring in the steady state. On the other hand, a *pure* spin current is free from such a restriction and can flow into the ring if the charge current $I$ is spin-polarized. Since this geometry is similar to the nonlocal spin valve geometry considered in Refs. [20,21] and the calculation is also similar to the nonlocal spin valve geometry case, we skip minor details of calculation and present the main results only. By solving appropriate diffusion equations for the entire system (ring and "spin injector"), one obtains the relation between the charge current $I$ and the pure spin current $I_s$ injected into the ring,

$$I_s = 2I \times \left[ \frac{\left(\frac{2p}{1-p^2}\frac{R_i}{R_N} + \frac{2\alpha}{1-\alpha^2}\frac{R_F}{R_N}\right)\exp\left(-\frac{L_{\text{source}}}{\lambda_N}\right)}{\left(\frac{2}{1-p^2}\frac{R_i}{R_N} + \frac{2}{1-\alpha^2}\frac{R_F}{R_N} + 1\right)\left(\frac{R_S}{R_N} + 2\right) + \left(\frac{R_S}{R_N} - 2\right)\exp\left(-\frac{2L_{\text{source}}}{\lambda_N}\right)} \right], \quad (B.1)$$

where the ferromagnetic electrode $F_s$ is assumed to be connected to the normal metal via a tunneling barrier with the same tunneling resistance $R_i$ as the tunneling resistance of the ring junctions. The result for the ohmic interface can be obtained from Eq. (B.1) by taking the limit $R_i \to 0$. Here $L_{\text{source}}$ is distance between $F_s$ and FI, and $R_s$ is given by

$$R_s = R_N \cdot \left[ C_3 + \frac{\frac{R_{\text{ring}}^{\text{ohm}}}{R_N} \cdot C_4 + \frac{2}{1-p^2}\frac{R_i}{R_N} \cdot C_5}{\frac{R_{\text{ring}}^{\text{ohm}}}{R_N} C_1 + \frac{2}{1-p^2}\frac{R_i}{R_N} C_2 + \frac{2\alpha^2}{1-\alpha^2}\frac{R_F}{R_N}(e^{\zeta_F}-1)(e^{\zeta_N}-1)} \right], \quad (B.2)$$

where the constants C1 and C2 are defined in Eq. (15) and the dimensionless constants C3, C4, and C5 are given by

$$C_3 = \frac{\frac{1}{1-p^2}\frac{R_i}{R_N}(e^{\zeta_F}-1)(e^{\zeta_N}+1) + \frac{1}{1-\alpha^2}\frac{R_F}{R_N}(e^{\zeta_F}+1)(e^{\zeta_N}+1)}{\frac{1}{1-p^2}\frac{R_i}{R_N}(e^{\zeta_F}-1)(e^{\zeta_N}-1) + \frac{1}{1-\alpha^2}\frac{R_F}{R_N}(e^{\zeta_F}+1)(e^{\zeta_N}-1) + (e^{\zeta_F}-1)(e^{\zeta_N}+1)},$$

$$C_4 = \frac{1}{1-p^2}\frac{R_i}{R_N}(e^{\zeta_F}+1)(e^{\zeta_N}-1) + \frac{1}{1-\alpha^2}\frac{R_F}{R_N}(e^{\zeta_F}-1)(e^{\zeta_N}-1),$$

$$C_5 = \frac{R_i}{R_N}(e^{\zeta_F}+1)(e^{\zeta_N}-1) + \frac{(1-2p\alpha+\alpha^2)}{1-\alpha^2}\frac{R_F}{R_N}(e^{\zeta_F}-1)(e^{\zeta_N}-1).$$

Below we consider a few limiting cases. When the length of paramagnetic and ferromagnetic metals of the ring are much longer than their spin diffusion length ($L_F, L_N \gg \lambda_F, \lambda_N$) so that $R_{\text{ring}}^{\text{ohm}} \gg R_F, R_N$, we obtain the following results depending on $R_i$; ( i ) If all the FN interfaces are ohmic so that $R_i \to 0$ the result for the spin current $I_s$ is

$$I_s = I \cdot \frac{\dfrac{2\alpha}{1-\alpha^2}\dfrac{R_F}{R_N}\left(1+\dfrac{1}{1-\alpha^2}\dfrac{R_F}{R_N}\right)\exp\left(-\dfrac{L_{\text{source}}}{\lambda_N}\right)}{\left(1+\dfrac{2R_F}{(1-\alpha^2)R_N}\right)^2 - \exp\left(-\dfrac{2L_{\text{source}}}{\lambda_N}\right)}, \qquad (B.3)$$

and ( ii ) if all the F/N interfaces are strong tunnel barrier contacts so that $R_i \gg R_N, R_F$ but $R_{\text{ring}}^{\text{ohm}} \gg R_i$,

$$I_s = I \cdot \frac{p}{2}\exp\left(-\frac{L_{\text{source}}}{\lambda_N}\right). \qquad (B.4)$$

Note that in the large ring limit, the spin current $I_s$ is still injected to the ring so long as the distance $L_{\text{source}}$ between the F$_s$/P$_s$ interface and the ring is sufficiently smaller than the spin flip length $\lambda_N$ of the P$_0$. Finally, when we revisit the limiting case $L_F, L_N \gg \lambda_F, \lambda_N$ and present the voltage difference $V_{\text{I-1}}$ at the interfaces I-1 and $V_{\text{I-2}}$ at the I-2 defined in Eqs. (9) and (10), we find for the case ( i ),

$$V_{\text{I-1}} \to \left[\frac{2\alpha^2}{(1-\alpha^2)^2}R_N\left(\frac{R_F}{R_N}\right)^2 \frac{\exp\left(-\dfrac{L_{\text{source}}}{\lambda_N}\right)}{\left(1+\dfrac{2R_F}{(1-\alpha^2)R_N}\right)^2 - \exp\left(-\dfrac{2L_{\text{source}}}{\lambda_N}\right)}\right] \times I, \qquad (B.5)$$

and for the case ( ii )

$$V_{\text{I-1}} \to \left[\frac{p^2}{2}R_N\exp\left(-\frac{L_{\text{source}}}{\lambda_N}\right)\right] \times I. \qquad (B.6)$$

For both case

$$V_{\text{I-2}} \to 0. \qquad (B.7)$$

Note that in this limit we recover the results of the spin valve system of Ref.[21] for $V_{\text{I-1}}$ corresponding to the spin signal injected to the interface I-1. Furthermore, by combining Eq. (B.3) with Eq. (8) or Eq. (B.4) with Eq. (16), one can also see that the voltage difference $V_{\text{I-1}}$ due to the spin accumulation plays a role in generating the charge current in the ring [see Eq. (11)].

## References

[1] G. Prinz, Science **282**, 1660 (1998); S. A. Wolf, D. D. Awschalom, R. A. Buhrman, J. M. Daughton, S. von Molna, M. L. Roukes, A. Y. Chtchelkanova, D. M. Treger, Science **294**, 1488 (2001).

[2] Q.-F. Sun, H. Guo, and J. Wang, Phys. Rev. Lett. **90**, 258301 (2003); W. Long, Q.-F. Sun, H. Guo, and J. Wang, Appl. Phys. Lett. **83**, 1397 (2003); S. K. Watson, R. M. Potok, C. M. Marcus, and V. Umansky, Phys. Rev. Lett. **91**, 258301 (2003).

[3] See for instance *Introduction to mesoscopic physics* by Y. Imry (Oxford University Press, New York, 1997).

[4] J. E. Hirsch, Phys. Rev. Lett. **83**, 1834 (1999).

[5] S. Murakami, N. Nagaosa, and S.-C. Zhang, Science **301**, 1348 (2003); J. Sinova, D. Culcer, Q. Niu, N. A. Sinitsyn, T. Jungwirth, and A.H. MacDonald, Phys. Rev. Lett. **92**, 126603 (2004).

[6] J. E. Hirsch, Phys. Rev. B **42**, 4774 (1990); *ibid.*, **60**, 14787 (1999).

[7] Q.-F. Sun, H. Guo, and J. Wang, Phys. Rev. B **69**, 054409 (2004).

[8] T. P. Pareek, Phys. Rev. Lett. **92**, 076601 (2004).

[9] A. Brataas, Y. Tserkovnyak, G. E. W. Bauer, and B. I. Halperin, Phys. Rev. B **66**, 060404(R) (2002).

[10] M. J. Stevens, A. L. Smirl, R. D. R. Bhat, A. Najmaie, J. E. Sipe, and H. M. van Driel, Phys. Rev. Lett. **90**, 136603 (2003).

[11] A. Fert and I. A. Campbell, J. Phys. (Paris) Colloq. **32**, C1-46 (1971); P. C. van Son, H. van Kempen, and P. Wyder, Phys. Rev. Lett. **58**, 2271 (1987).

[12] T. Valet and A. Fert, Phys. Rev. B **48**, 7099 (1993).

[13] S. Hershfield and H. L. Zhao, Phys. Rev. B **56**, 3296 (1997).

[14] F. J. Jedema, M. S. Nijboer, A. T. Filip, and B. J. van Wees, Phys. Rev. B **67**, 085319 (2003).

[15] *Semiconductor Spintronics and Quantum Computation*, edited by D. D. Awschalom, D. Loss, and N. Samart (Springer-Verlag, Berlin, 2002).

[16] E. I. Rashba, Phys. Rev. B **62**, R16267 (2000).

[17] M. Johnson, Phys. Rev. Lett. **70**, 2142 (1993).

[18] A. Fert and S.-F. Lee, Phys. Rev. B **53**, 6554 (1996).

[19] M. Johnson and R.H. Silsbee, Phys. Rev. B **37**, 5312 (1988).

[20] A. T. Filip, B. H. Hoving, F. J. Jedema, B. J. van Wees, B. Dutta, and S. Borghs, Phys. Rev. B **62**, 9996 (2000); F. J. Jedema *et al.*, Nature **416**, 713 (2002).

[21] S. Takahashi and S. Maekawa, Phys. Rev. B **67**, 052409 (2003).

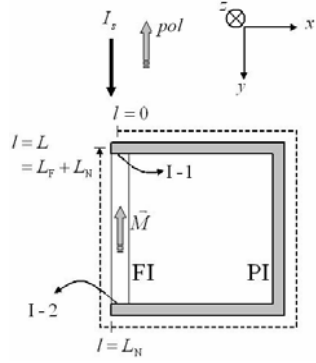

Fig. 1: Schematic diagram of a ferromagnet(F)-normal metal(N) hybrid ring. A pure spin current $I_s$ is injected to the ring at $l=0^+$, that is, the normal metal side of the interface I-1. The grey region ($0<l<L_N$) represents the normal metal with length $L_N$ while the white region ($L_N<l<L=L_N+L_F$) the ferromagnet with length $L_F$. Both the magnetization($\vec{M}$) of the F and the spin polarization($pol$) of the injected spin current are along -y axis.

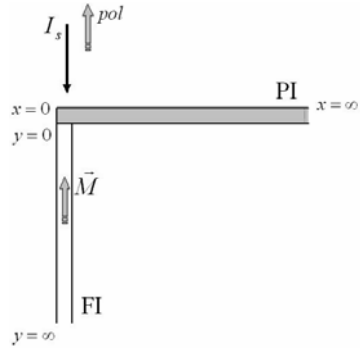

Fig. 2: A geometry obtained by deforming the configuration in Fig. 1 in the large ring limit.

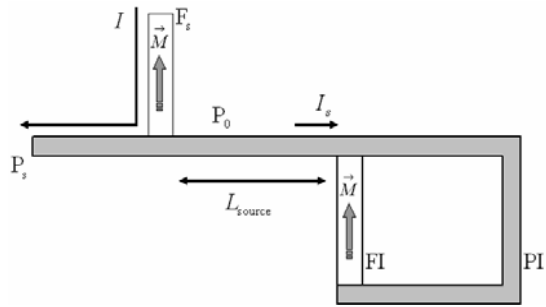

Fig. 3: A particular geometry for the injection of a pure spin current into the ring. This geometry is similar to the nonlocal spin-valve geometry in Refs. [20, 21].